\documentclass[rnote]{aa} % for the research notes
\usepackage{longtable} 
\usepackage{graphicx}
\usepackage{graphics}
\usepackage{natbib}
\usepackage{amsmath}
\usepackage{subfig}
\def \psr{PSR\, B0656+14}

\newcommand{\hst}{{\sl HST}}

\newcommand{\chan}{{\sl Chandra}}

\begin{document}

\title{VLT polarimetry observations of PSR\, B0656+14\thanks{Based on observations collected at ESO, Paranal, under Programme 090.D-0106(A)}}
\author{R. P. Mignani\inst{1,2}
\and
P.~Moran\inst{3}
\and
A.~Shearer\inst{3}
\and
V.~ Testa\inst{4}
\and
A.~S\l{}owikowska \inst{2}
\and
B.~Rudak\inst{5}
\and
K.~Krzeszowki\inst{2}
\and
G.~Kanbach\inst{6}
}

 \institute {INAF - Istituto di Astrofisica Spaziale e Fisica Cosmica Milano, via E. Bassini 15, 20133, Milano, Italy
  \and Kepler Institute of Astronomy, University of Zielona G\'ora, Lubuska 2, 65-265, Zielona G\'ora, Poland
  \and Centre for Astronomy, School of Physics, National University of Ireland Galway, University Road, Galway, Ireland
  \and INAF - Osservatorio Astronomico di Roma, via Frascati 33, 00040, Monteporzio, Italy 
  \and Nicolaus Copernicus Astronomical Center, ul. Rabia\'nska 8, 87100, Toru\'n, Poland 
   \and Max-Planck Institut f\"ur Extraterrestrische Physik, Giessenbachstrasse 1, 85741 Garching bei M\"unchen, Germany 
        }
        
\titlerunning{Optical polarimetry of PSR\, B0656+14}
\authorrunning{Mignani et al.}
\offprints{R. P. Mignani; mignani@iasf-milano.inaf.it}
\date{Received ...; accepted ...}
\abstract{Optical  polarisation measurements are key tests for different models  of the  pulsar magnetosphere. Furthermore, comparing the relative orientation of  the phase-averaged linear polarisation direction and the pulsar proper  motion vector may unveil a peculiar alignment, clearly seen in the Crab pulsar.  
}
{Our goal is to obtain the first measurement of the phase-averaged optical linear polarisation of the fifth brightest optical pulsar, PSR\, B0656+14, which has also a precisely measured proper motion, and verify a possible alignment between the polarisation direction and the proper motion vector.}
{We carried out observations with the Very Large Telescope (VLT) to measure the phase-averaged optical polarisation degree (P.D.)  and position angle (P.A.) of \psr.}
{We measured a P.D. of $11.9\%\pm5.5\%$ and a P.A. of $125.8\degr\pm13.2\degr$, measured East of North.
Albeit of marginal significance, this is the first measurement of the phase-averaged optical P. D. for this pulsar. 
Moreover, we found that the P.A. of the phase-averaged polarisation vector is 
close to that of the pulsar proper motion  ($93.12\degr\pm0.38\degr$).
 }
{Deeper observations are needed to confirm our polarisation measurement of \psr, whereas polarisation measurements for more pulsars will better assess possible correlations of the polarisation degree with the pulsar parameters.}

 \maketitle

\section{Introduction}

Polarisation measurements of pulsars offer unique insights into  their highly-magnetised relativistic  environments  and are a primary test for neutron  star magnetosphere models and radiation emission mechanisms. Besides the radio band,  optical observations are best suited to these goals.
Indeed, significant polarisation is expected when the optical emission is  produced by synchrotron  radiation in the neutron star magnetosphere.  Owing to their faintness linear optical polarisation measurements exist for a handful of  pulsars only,  whereas the circular polarisation was measured only for  the Crab pulsar  (Wiktorowicz et al.\ 2015).
The young ($\sim960$ year old) and bright ($V\sim 16.5$) Crab pulsar (PSR\, B0531+21) is 
 the only one for which precise measurements  have been repeatedly obtained, both phase-resolved  (e.g., S\l{}owikowska et al.\ 2009) and phase-averaged (e.g., Moran et al.\ 2013).  Interestingly, the phase-averaged optical polarisation position angle is similar to that of  the pulsar  proper motion direction and close to that of the axis of  symmetry of the  X-ray arcs and jets  of the pulsar-wind nebula (PWN) observed by \chan\ (e.g. Hester 2008), which is assumed to be  also the pulsar  spin axis.  Such an approximate alignment,  might trace the connection between the  pulsar  magnetospheric  emission, its  dynamical  interaction with the PWN, and the neutron star formation, with the kick and spin axis directions determined at birth. 
Evidence of a similar alignment has been found from optical polarisation measurements of the Vela  pulsar (PSR\, B0833$-$45; $V\sim 23.6$) by Mignani et  al.\ (2007) and Moran et al.\ (2014), whereas for both PSR\, B0540$-$69  ($V=22.5$) and  PSR\, B1509$-$58 (R=25.7) the lack of a measured proper motion makes it impossible to check the alignment with the polarisation direction (Mignani et al.\ 2010; Lundqvist et al.\ 2011; Wagner \& Seifert 2000). 
The next best target 
is PSR\, B0656+14, the brightest pulsar detected in the optical after Vela (Mignani\ 2011).  
PSR\, B0656+14 ($V\sim 25$) is  a middle-aged ($\sim 10^{5}$ years) optical pulsar  (Shearer et al.\ 1997; Kern et  al.\  2003; Kargaltsev \& Pavlov 2007) with strong magnetospheric optical emission  (e.g., Zharikov et  al.\  2007). 
It is not embedded in a bright and variable  optical PWN and its supernova remnant (SNR) has already faded away in the interstellar medium (ISM), which
minimises the local background polarisation.
Moreover,  at a distance of $0.28\pm0.03$  kpc
(Verbiest et al.\ 2012),  PSR\, B0656+14 is affected by a relatively low foreground polarisation.  Last but not least, it has an extremely accurate radio proper motion ($\mu=44.13\pm0.63$ mas~yr$^{-1}$; Brisken et al.\  2003),
which makes it possible  to search for a possible alignment with the phase-averaged  polarisation direction.  

Here, we report on the first phase-averaged linear polarisation observations of PSR\, B0656+14, performed with the Very Large Telescope (VLT).

\section{Observations and Data  Analysis}

We observed PSR\, B0656+14 between January and February 2014 using the second version of the Focal Reducer and low dispersion Spectrograph (FORS; Appenzeller et al.\ 1998)  at the VLT Antu telescope.  
FORS2 is equipped with
a mosaic of two 4k$\times$2k MIT CCDs aligned along the Y axis, optimised for wavelengths longer than 6000 \AA. 
Observations were performed in  imaging polarimetry mode (IPOL), with standard low gain, normal readout (200 Kpix/s), standard-resolution mode  (0\farcs25/pixel), and the high efficiency v$_{\rm HIGH}$ filter  ($\lambda=555.0$ nm, $\Delta\lambda=61.6$ nm). For all the observations, the target was
at the nominal aim position on the upper CCD chip (CHIP1), to avoid the effects of instrumental polarisation at the CCD edges. 
The FORS2 polarisation optics consist of a Wollaston prism as a beam splitting analyser and two super-achromatic phase  retarder
3$\times$3 plate  mosaics 
 installed on rotatable mountings 
 that can be moved in and out of the light path.
  Images are obtained by taking two frames displaced by 22$^{\prime\prime}$ in the direction perpendicular to the Multi-Object Spectrograph  (MOS) slitlets. We used the four standard IPOL 
half-wave retarder plate angles of 0$^\circ$, 22.5$^\circ$, 45$^\circ$, and 67.5$^\circ$,
corresponding to the retarder plate orientation with respect  to the Wollaston prism and usually set with an accuracy $\la$0.1\degr\ (Boffin 2014).
Both the axis of the detector optics and the zero point of the half wave retarder plate angle  are set such that the polarisation position angle is measured eastward from North.

%_________________________________________________________________
%
\begin{figure}
\centering
\includegraphics[height=6.0cm]{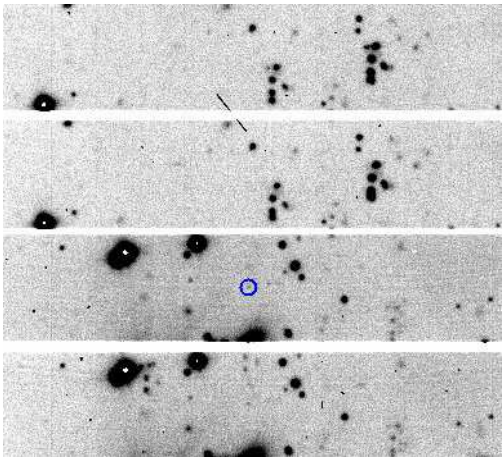}
\caption{
Image of the \psr\  field (715 s)
taken with the 
$0\degr$ retarder plate angle. 
The pair of images (22\arcsec\ height) correspond to the extraordinary and ordinary beams. The pulsar counterpart is circled. }
\label{fig4}
\end{figure}
%_________________________________________________________________
%

In order to not  introduce spurious effects 
due to the variable sky polarisation background across different nights, each of the nine observation blocks (OBs) incorporated exposures for all the four retarder plate angles (715 s each).
Observations were executed 
in dark time, with a seeing of 0\farcs6--1\farcs3, airmass below 1.4, and in clear sky conditions.  
Short exposures ($<$3 s) of both polarised
and unpolarised
standard stars were also acquired for calibration. Twilight flat-field images with no retarder plate along the light path were also acquired on the same nights as the science images.
For both the pulsar and standard stars, single-exposure raw images were bias-subtracted and flat-fielded using standard routines in {\sc IRAF}\footnote{IRAF is distributed by the National Optical Astronomy Observatories, which are operated by the Association of Universities for Research in Astronomy, Inc., under cooperative agreement with the National Science Foundation.}.  We computed the FORS2 astrometry 
using stars from the Guide Star Catalogue 2 (GSC-2; Lasker et al.\ 2008) as a reference. After accounting for all uncertainties (star centroids, GSC-2 absolute coordinate accuracy) our astrometric solution turned out to be  $\la$0\farcs1 rms. 
\psr\ (Fig.\ 1)  is detected at the position expected by extrapolating its proper motion (Brisken et al.\ 2013)  to the epoch of our observations, where the uncertainty on the proper motion extrapolation is well below that on our astrometry calibration.

In order to increase the signal--to--noise, we co-added all the nine reduced science images of \psr\ taken with the same retarder plate angle.
For each angle, we aligned the single images using the {\sc IRAF} tasks {\tt ccdmap} and {\tt ccdtrans}, with an average accuracy of a few hundreds of a pixel.  We applied the image co-addition with the routine {\tt combine} and filtered cosmic ray hits  using the {\em pclip} algorithm. Then,  we measured the pulsar flux in each of the four co-added image 
through PSF photometry using the package 
{\tt daophot} (Stetson 1994)
 implemented in {\sc IRAF}.  
In particular, we  fitted the model PSF to the pulsar intensity profile within an area of 10 pixel radius (2\farcs5),
estimated from the growth curves of reference stars. 
We measured the sky background in an annulus of inner radius of  10.5 pixels and width of 10 pixels (2\farcs6 and 2\farcs5), respectively, centred on the pulsar position. We carefully tailored 
this annulus both to avoid the wings of the pulsar intensity profile and accurately sample the sky background without being sensitive to strong fluctuations and gradients  in the background level.  
We  corrected the pulsar flux for the atmospheric extinction using, for each image, the average airmass value and  extinction coefficients in the $v_{\rm HIGH}$ filter\footnote{{\tt www.eso.org/qc}}.
We followed 
Fossati et al.\ (2007) to compute the errors in P.D. and P.A. 
From the observations of the polarised standards, 
the absolute calibration of our polarimetry is accurate to $\sim 0.1\%$ in P.D. and to $\la0.5$\degr\ in P.A., whereas from the observations of the unpolarised standards we found 
no evidence of significant instrumental polarisation at the CHIP1 aim position
and no systematic deviations in the zero point of the retarder half wave plate angle. 
For the photometry 
parameters defined above, we measured a $\rm P.D.=11.9\% \pm5.5\%$  and a  $\rm P.A.=125.8\degr\pm13.2\degr$, where the associated errors account both for statistical errors and calibration uncertainties. 

\section{Discussion}

Although of marginal significance ($\sim 2.2 \sigma$), but still comparable to those obtained for other pulsars (Mignani et al.\ 2007; Lundqvist et al.\ 2011), ours is the first and only  measurement of the phase-averaged optical polarisation degree of \psr\ obtained so far.  Indeed, no value of the phase-averaged polarisation was reported from the phase-resolved polarimetry observations of Kern et al.\ (2003). 
Their measurements 
show that the bridge between the two peaks of the optical light curve seems to be strongly polarised (P.D. $\approx 100\%$), whereas the peaks seem to be unpolarised (P.D. $\approx 0\%$). However, the large errors ($\pm 40\%$ at $1 \sigma$) attached to the phase-resolved polarisation values in each phase bin strongly affect the significance of their result.
Nonetheless, such a large variation 
of the P.D.  
would suggest that the phase-average value obtained from their data would be quite low and, possibly, close to our value. 
 
 \begin{table*} 
  \begin{center}
  \caption{Summary of the phase-average linear polarisation measurements for pulsars detected in the optical.   For the Crab,  
   a value of the  polarisation was obtained  from phase-resolved observations by S\l{}owikowska et al.\ (2009; 2012).
}
\label{sum}
 \begin{tabular}{lcccccccc}
 \hline
 Pulsar & $\tau$ & $P_{\rm s}$ & $\dot{P}_{\rm s}$ &  $\dot{E}$ & $B_{\rm S}$ & $B_{\rm LC}$ & P.D. & References \\ 
             & ($10^{3}$  yr)    &  (s)               & ($10^{-13}$ s s$^{-1}$)         & ($10^{38}$ erg cm$^{-2}$ s$^{-1}$)  & ($10^{12}$  G)  & ($10^{5}$ G) & (\%)& \\ \hline
B0531+21	            &    1.24  & 0.033	       &  4.22 	&    4.6  	&  3.78      &      9.80 	& 	5.2$\pm$0.3 & (1)  \\ 
                             &              &                     &               &               &               &                      &       5.5$\pm$0.1                & (2) \\
B0540$-$69	&  1.67  & 0.050	& 	4.79 	       	&  1.5  &  4.98          & 3.62	      &   5.0$\pm$2.0 & (3)  \\
      	 &                        &             &                                               &                          &                                   &                              &  16.0$\pm$4.0& (4) \\
	   	 &                        &             &                                               &                          &                                   &                              &  $\approx$5.0&  (5) \\
B1509$-$58	 &  1.56 	 & 0.151	& 	15.3        &  0.17   & 15.40       	&  0.42          &  10.4 & (5) \\
B0833$-$45  & 11.3 	& 0.089	      &   1.25 	    & 	0.069   	&  3.38  	&  0.44 	& 	8.1$\pm$0.7 &  (6) \\
  &  	&	      &   	    & 	  	&   	&  	& 	9.4$\pm$4 &  (7)  \\
  &  	&	      &   	    & 	  	&   	&  	& 	8.5$\pm$0.8 &  (5) \\
B0656+14	      &    111   & 0.384	       &  0.55	 & 	0.00038       &  4.66   	 &  0.007 		& 11.9$\pm$5.5 & this work \\
 \hline
 \end{tabular}
 \end{center}
(1) Moran et al.\ (2013); (2) S\l{}owikowska et al.\ (2012); (3)  Lundqvist et al.\ (2011); (4) Mignani et al.\ (2010); (5) Wagner \& Seifert (2000); (6) Moran et al.\ (2014);  (7) Mignani et al.\ (2007)
\end{table*}

We also obtained a first measurement of the P.A. of the phase-averaged optical polarisation of  \psr.
The 
value ($125.8\degr\pm13.2\degr$) 
is close to the P.A. of  the pulsar proper motion 
($93.12\degr \pm0.38\degr$; Brisken et al.\ 2003).
Since the difference between the two angles 
is within $3\sigma$, an alignment between the two vectors is possible,
as clearly seen in the Crab pulsar (Moran et al.\ 2013) and, approximately, also in the Vela pulsar (Moran et al.\ 2014). 
Evidence of alignment between the polarisation and proper motion vectors is also found in radio
(Johnston et al.\  2005; Noutsos et al.\ 2012).
For \psr, radio polarisation measurements at 3.1 GHz  give a P.A. of  $-86\degr \pm2\degr$ (Johnston et al.\ 2007) at the phase of closest approach of the magnetic pole to the observer's line of sight. This value is very close (mod 180\degr) to the proper motion P.A. and, 
accounting for 
the phase-dependent variations
of the radio polarisation P.A. 
and the measurement uncertainties,
is comparable to our phase-averaged optical polarisation P. A.
Assuming that the pulsar proper motion vector  is always aligned with the spin axis (e.g., Lai et al.\ 2001; Johnston et al.\ 2005, 2007; Noutsos et al.\ 2012), the alignment with the optical polarisation direction might be explained by a similar optical emission geometry and inclination angle $\alpha$ of the magnetic axis with respect to the spin axis for the Crab and Vela pulsars and \psr. This is  suggested by 
%the fact that they all have 
their two-peak optical light curves and possibly similar values of $\alpha$ (e.g. Johnston et al.\ 2005; McDonald et al.\ 2011; 
%Du et al.\ 2011;  
Zhang \& Jiang 2006).  
For both the Crab and Vela pulsars, 
the phase-averaged optical polarisation and proper motion  vectors are also approximately aligned with the axis of symmetry of their X-ray PWNe, thought to coincide with the pulsar spin axis. Unfortunately, for \psr\ only a tentative evidence of an X-ray PWN has been found  so far (Pavlov et al.\ 2002).

\begin{figure*}
\centering
\begin{tabular}{cc}
{\includegraphics[width=85mm,bb=0 0 576 320,clip=]{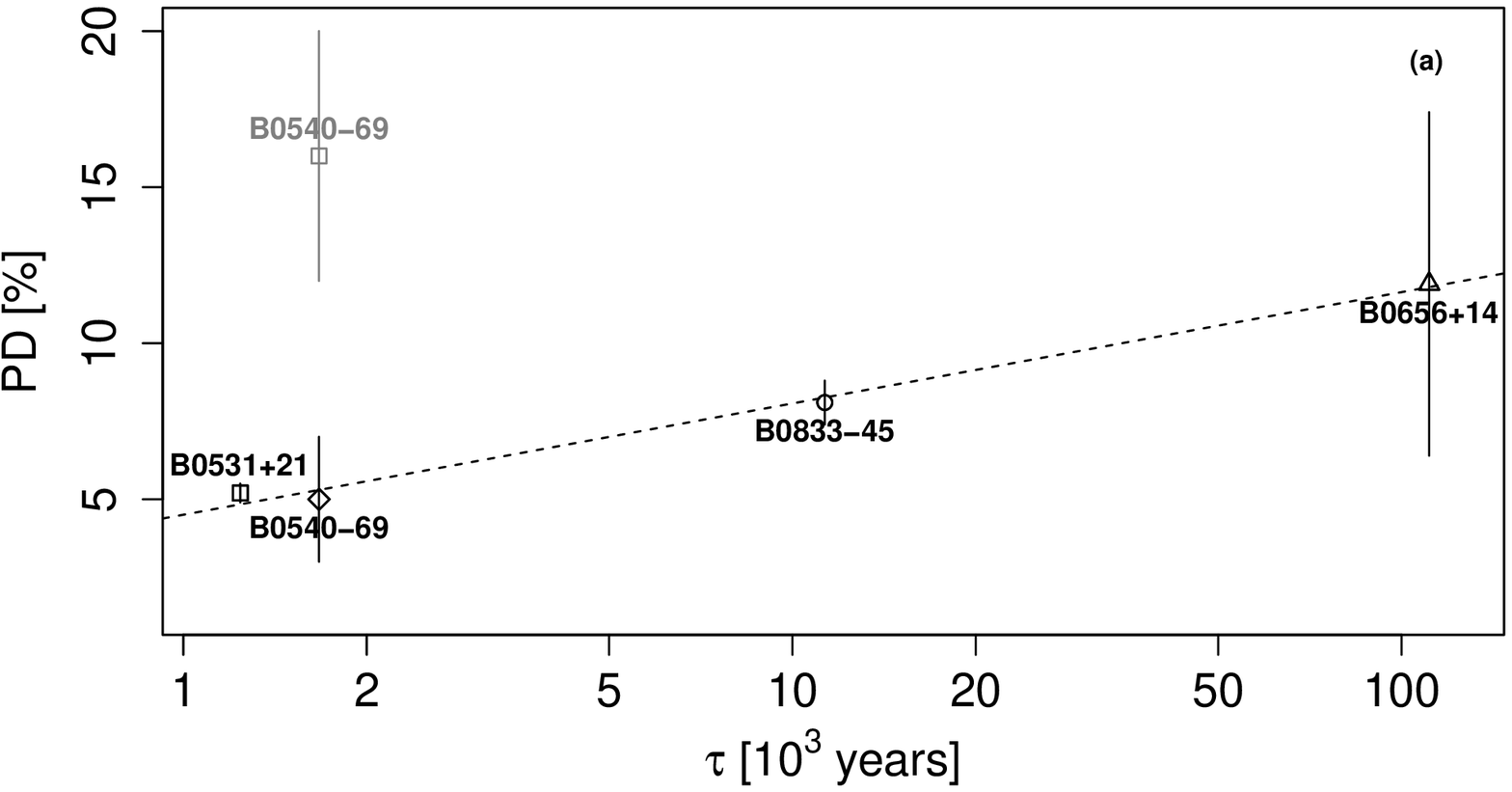}} &
{\includegraphics[width=85mm,bb=0 0 576 320,clip=]{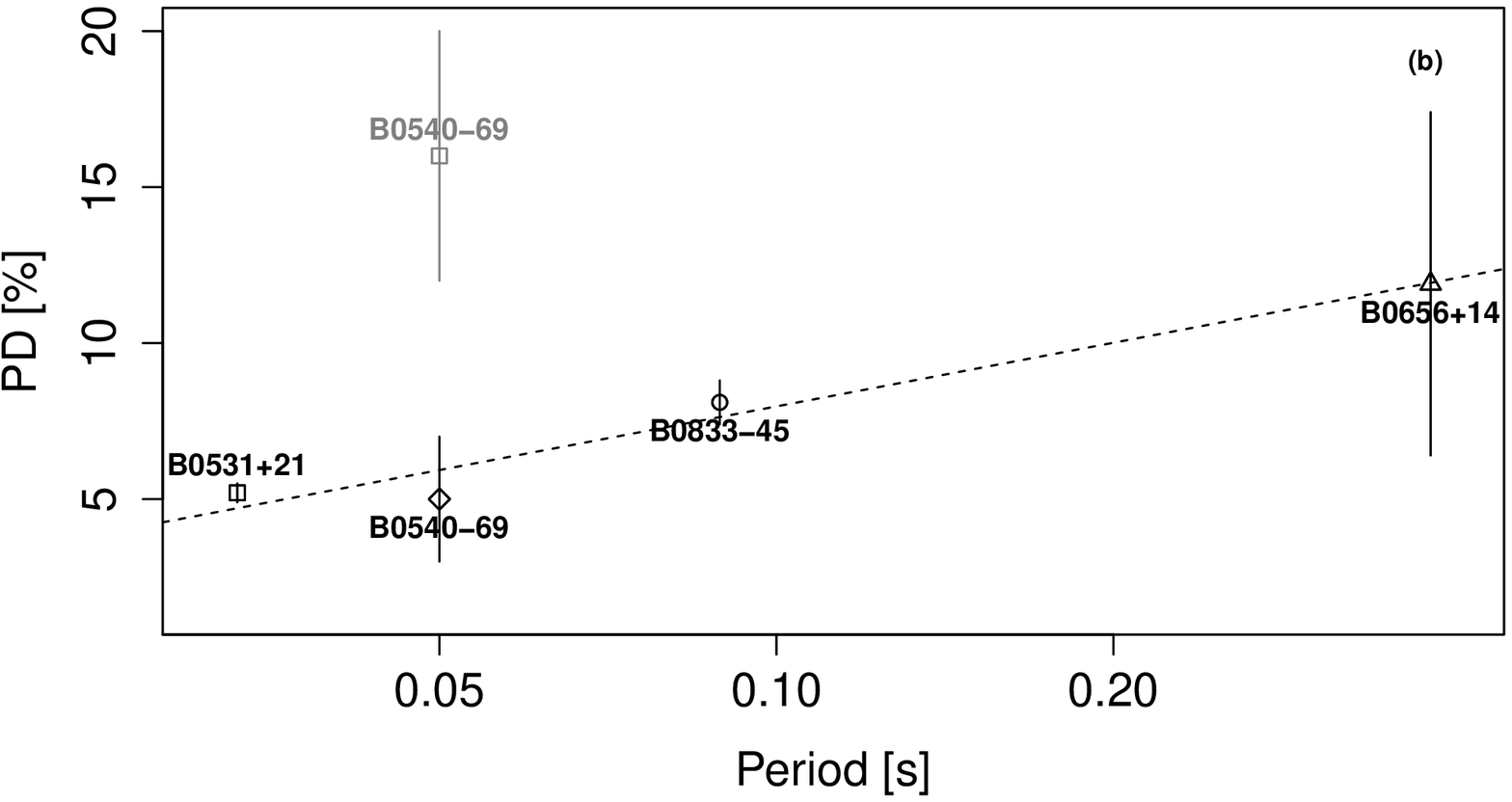}} \\
{\includegraphics[width=85mm,bb=0 0 576 320,clip=]{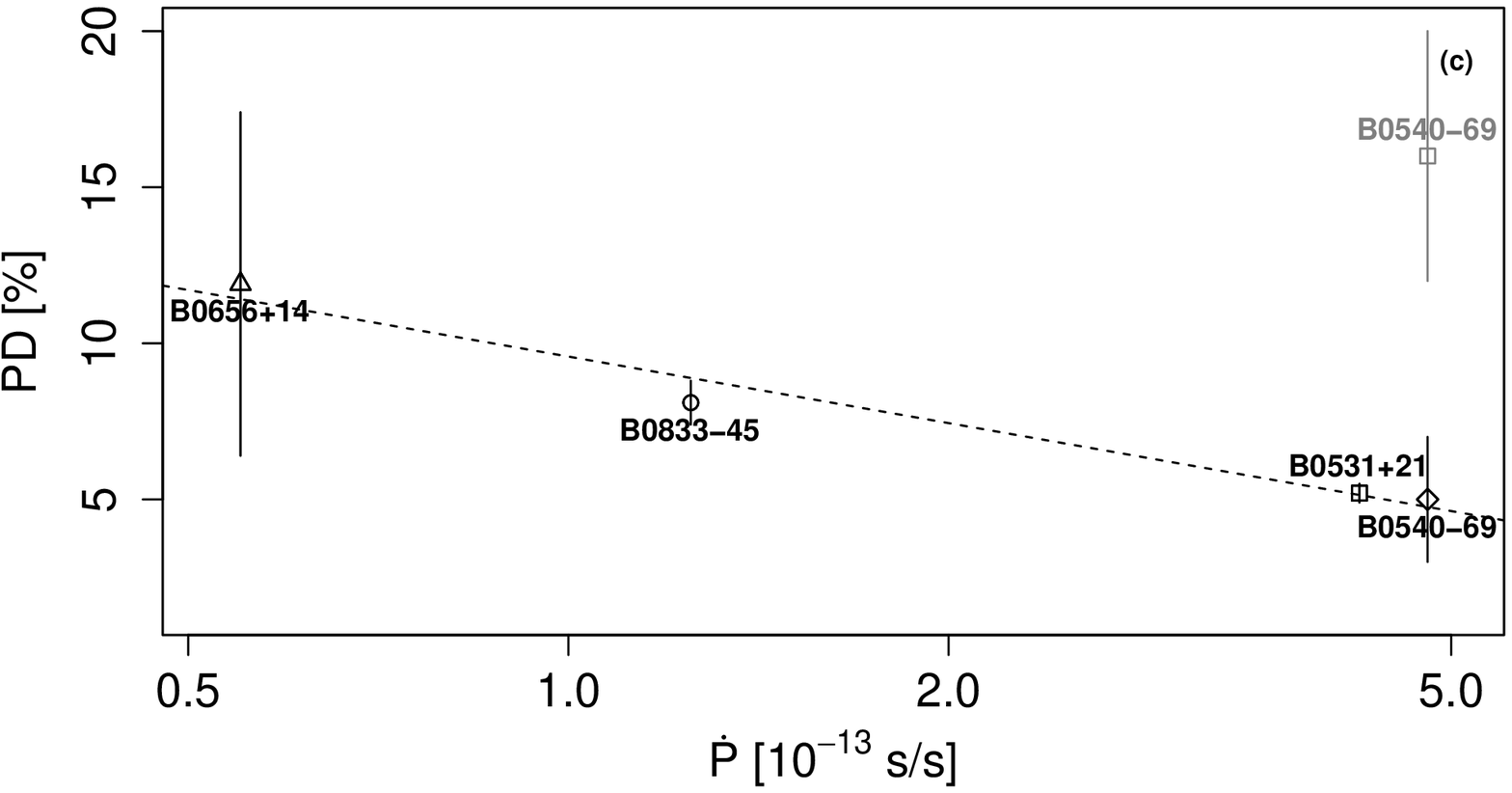}} &
{\includegraphics[width=85mm,bb=0 0 576 320,clip=]{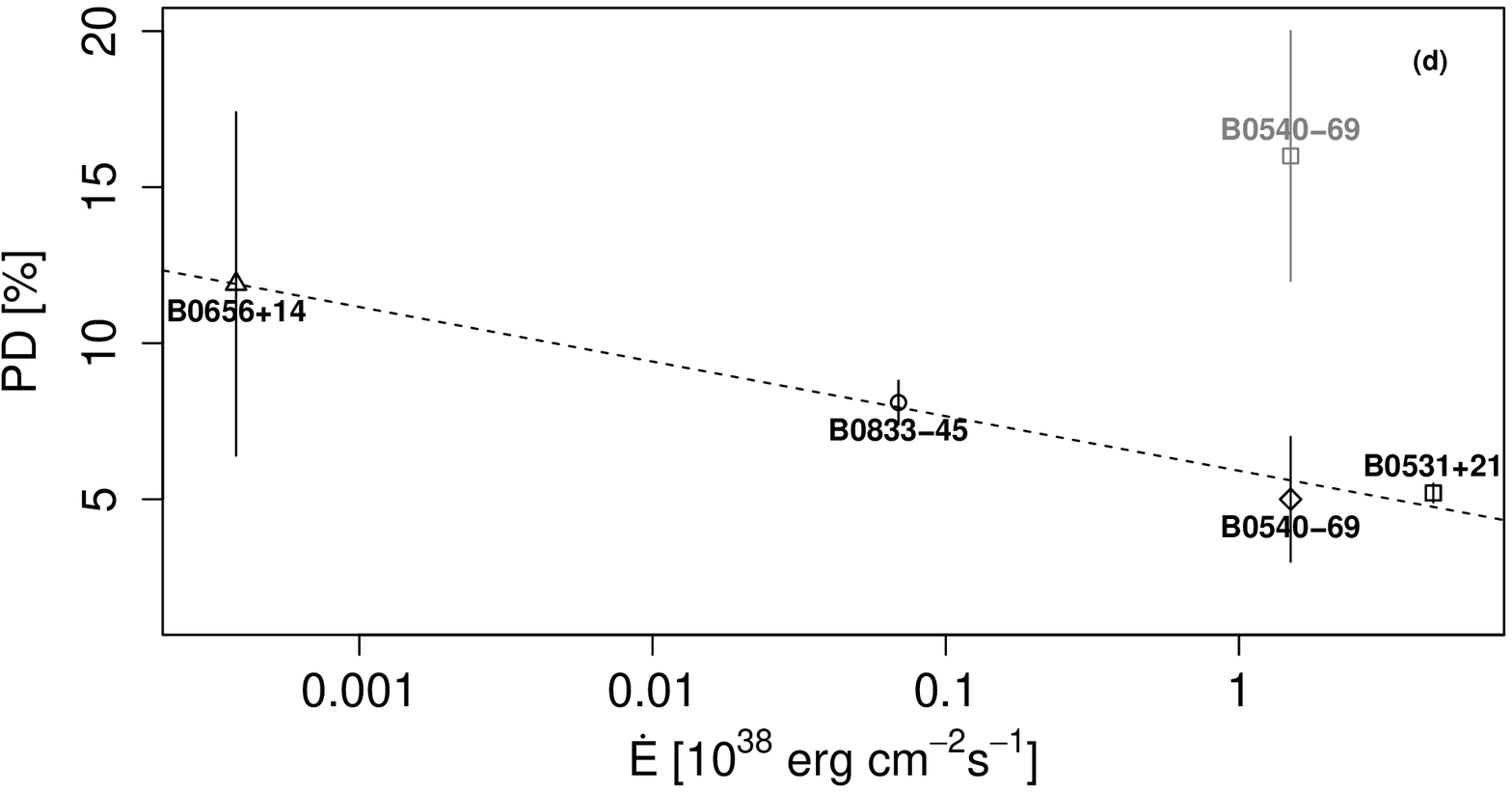}} \\
{\includegraphics[width=85mm,bb=0 0 576 320,clip=]{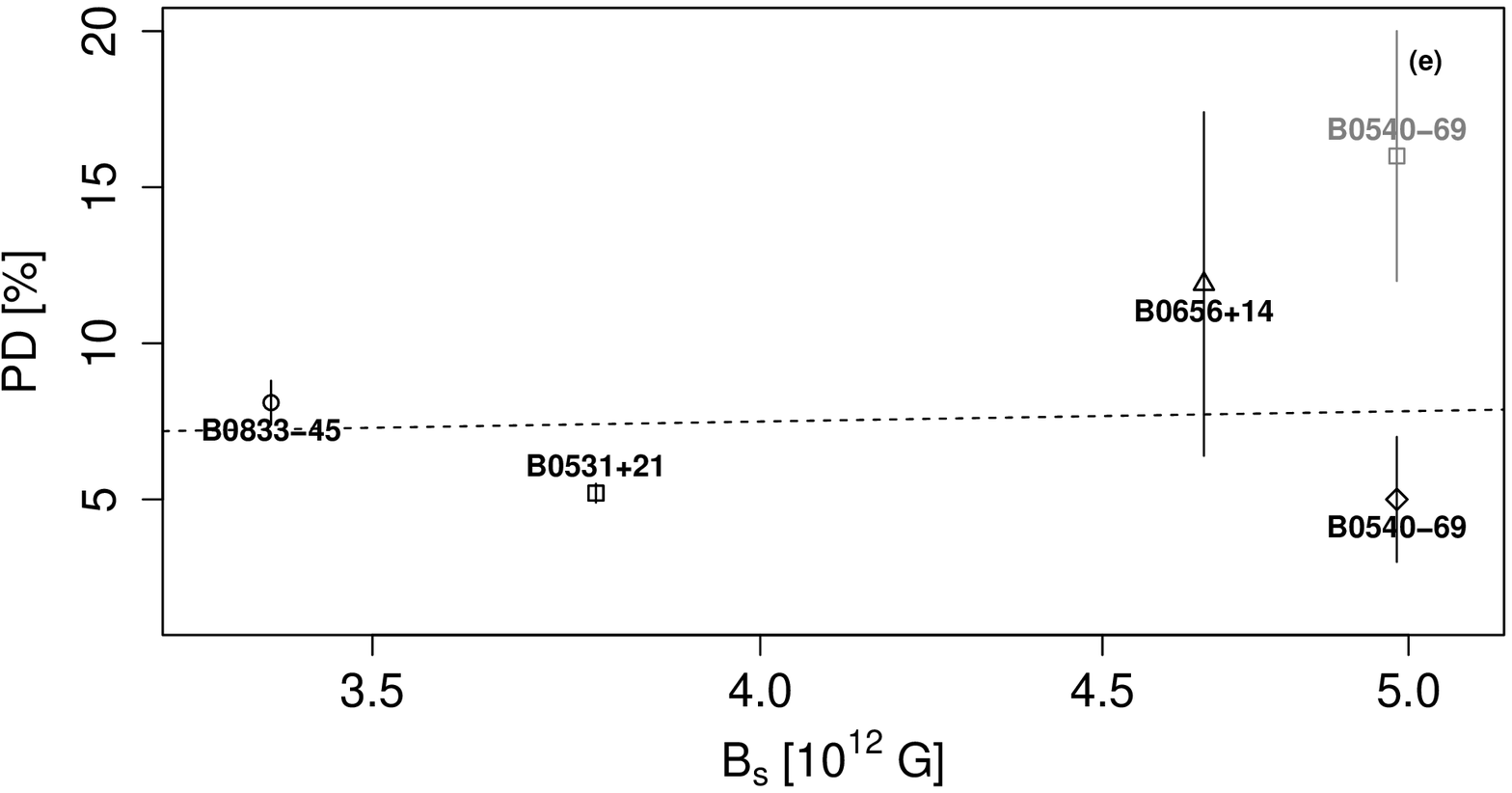}} &
{\includegraphics[width=85mm,bb=0 0 576 320,clip=]{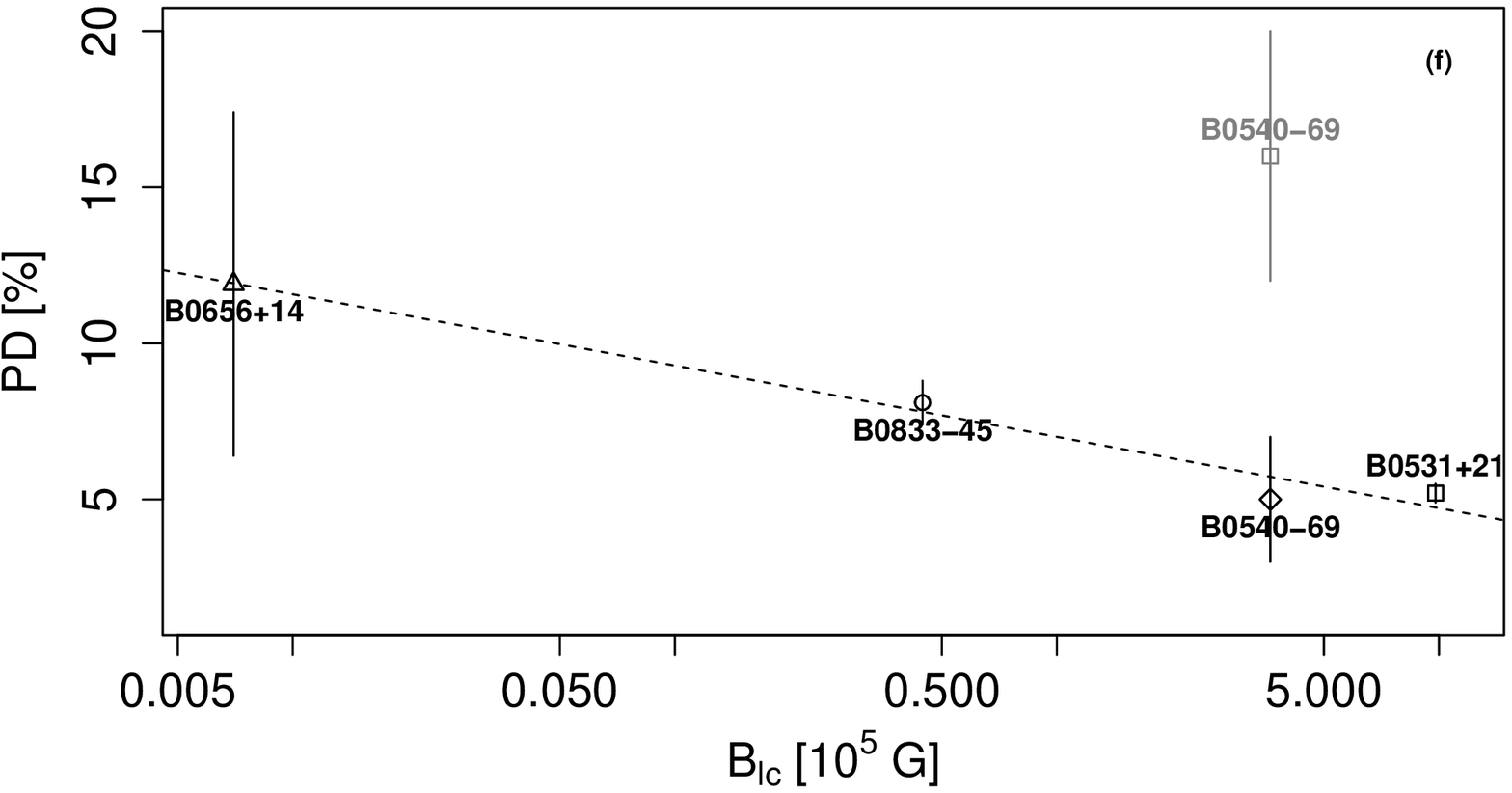}} \\
\end{tabular}
\caption{Plots of the P.D. 
as a function of a) the pulsar characteristic age $\tau$, b) spin period $P_{\rm s}$, c) spin period derivative $\dot{P}_{\rm s}$, d) spin-down power $\dot{E}$, e) magnetic field at the surface $B_{\rm s}$, and f)  at the light cylinder $B_{\rm LC}$. 
The dotted lines show the best fit to the points with a linear function. The point in light grey deviates more than $1 \sigma$ from the best fit. }
\label{figure2}
\end{figure*}

With our VLT measurement, 
linear polarisation values have now been obtained for all the five brightest pulsars identified in the optical (Mignani 2011).
Table \ref{sum} summarises all the
measurements obtained from  phase-averaged imaging polarimetry.
From a general standpoint, these data suggest that 
the pulsar phase-averaged polarisation is lower
in the optical than in radio (e.g., Weltevrede \& Johnston 2008).
This 
might be ascribed to the difference between incoherent and coherent radiation emission mechanisms in the optical and in radio, respectively. 
Such a comparison, however, must be taken with due care since the radio band is much broader than the optical one and the radio P.D. 
is frequency-dependent. In particular, it decreases from the MHz to the GHz 
frequency ranges, 
probably due to de-polarisation effects produced by scattering of the radio waves in the ISM (e.g., Noutsos et al.\ 2009).
X and $\gamma$-ray polarisation measurements could confirm that the polarisation level depends on the underlying emission mechanisms.
We 
investigated possible correlations between the phase-averaged optical P.D. and the pulsar characteristic age $\tau$, the spin period $P_{\rm s}$ and  its derivative $\dot{P}_{\rm s}$, the spin-down power $\dot{E}$, and the pulsar magnetic field measured both at the surface ($B_{\rm s}$) and at the light cylinder ($B_{\rm LC}$).  
For both the Crab and Vela pulsars, we assumed the recent \hst\ measurements (Moran et al.\ 2013; 2014) as a reference.  We did not include the VLT polarisation measurements of PSR\, B0540$-$69 and PSR\, B1509$-$58 (Wagner \& Seifert 2000), which have no associated errors. 
The linear fit shows some evidence of a possible increase of P.D.  with 
$\tau$ (Fig.\ 2a; reduced $\chi_{r}^2=0.38$).
 Possible trends for an increase of P.D. with  $P_{\rm s}$ 
  (Fig.\ 2b; $\chi_{r}^2=0.84$) and for a decrease of P.D. with 
  $\dot{P}_{\rm s}$ (Fig.\ 2c; $\chi_{r}^2=0.33$) are also found.
The data also suggest a possible decrease of P.D. with $\dot{E}$
(Fig.\ 2d; $\chi_{r}^2=0.58$). Incidentally, we note that the opposite trend
 is observed in  radio, as found, e.g. by observations at 4.9 GHz (von Hoensbroech et al.\ 1998) and 1.5 GHz (Weltevrede \& Johnston 2008), although such a trend is more clear at high radio frequencies.
Since the optical luminosity scales with $\dot{E}$ (e.g., Mignani et al.\ 2012), the trend in Fig.\ 2d would also imply that the fainter pulsars are more strongly polarised than the brighter ones. On the other hand, there is no correlation between the optical P.D. and $B_{\rm s}$
(Fig.\ 2e; $\chi_{r}^2=14.5$),
whereas a possible trend for a decrease  of P.D. with $B_{\rm LC}$
is more apparent (Fig.\ 2f; $\chi_{r}^2=0.65$).  
To summarise, if our measurement of the \psr\ polarisation were confirmed the general picture that would emerge 
would be that the P.D. tends to be higher in older and less energetic pulsars, and with the lower value of the magnetic field at the light cylinder.  The relatively lower values of P. D. for younger and more energetic pulsars might be due to the optical emission coming from spatially extended regions of the magnetosphere (and possibly forming caustic-like structures). In such a case, the resulting intrinsic P. D. would be lower due to effective depolarisation of the radiation from different emitting regions. The emitting regions (related to outer gaps or slot gaps) might shrink in radial extension as the pulsar ages and becomes less energetic. 
More optical polarisation measurements covering larger ranges in the parameter space,  together with the confirmation of the uncertain ones, will assess the reality of the observed trends and link them to the pulsar physical properties.  

\begin{acknowledgements}
This work is dedicated to the memory of our dear friend and colleague, Prof. Janusz Gil.
We are indebted to S. Bagnulo and A. Stinton for their help in the observations planning and data analysis.
RPM thanks the European Commission Seventh Framework Programme (FP7/2007-2013) for their support under grant agreement n.267251. 
PM is grateful for his funding from the Irish Research Council (IRC).  BR acknowledges support by the National Science Centre grant DEC- 2011/02/A/ST9/00256.
AS and KK acknowledge the Polish National Science Centre grant DEC-2011/03/D/ST9/00656.
We thank the referee for his/her constructive review of our manuscript.
\end{acknowledgements}

\end{document}